\documentclass[a4paper,12pt]{article}
\usepackage{amsmath,amssymb}
\begin{document}

\begin{center}

\textbf{THERMAL QRPA EQUATIONS WITH FINITE RANK SEPARABLE
APPROXIMATION FOR RESIDUAL FORCES BASED ON THE SKYRME TYPE
INTERACTIONS}

\bigskip
Alan A. Dzhioev$^{1,}$\footnote{On leave from Bogoliubov Laboratory of Theoretical Physics, Joint Institute
for Nuclear Research, Dubna, 141980 Russia},
 A. I. Vdovin$^2$

\medskip
$^1$Department of Physics, Universit\'e Libre de Bruxelles, Campus
Plaine, CP 231, Blvd du Triomphe, B-1050 Brussels, Belgium

\smallskip
$^2$Bogoliubov Laboratory of Theoretical Physics, Joint Institute
for Nuclear Research, Joliot-Curie, 6, Dubna, 141980 Russia

\end{center}

\noindent {\bf Abstract} The approach to study properties of charge-exchange excitations in hot nuclei is presented.
The approach is based on the extension of the finite rank separable approximation for Skyrme interactions to finite temperatures employing the TFD formalism. 
We present the formulae to obtain charge-exchange strength distributions within the Thermal Quasiparticle  Random Phase Approximation (TQRPA). 

\section{Introduction}

The properties of nuclei at finite temperatures appear to be
interesting for many reasons. Currently, one of the most popular
fields to apply the nuclear theory at finite temperatures is the
astrophysical studies. In the astrophysical context the theory is
used to calculate thermal modifications of spin-isospin transition
distributions over nuclear spectra, as they play an important role
in weak-interaction mediated reactions in stellar environment.

At first, temperature effects were introduced into the calculations
via quite straightforward way  taking into account a possibility of
thermal feeding of nuclear excited states which energies were
calculated via independent-particle shell model or taken from
experimental compilations \cite{Fuller80}. Later this approach was
developed and refined employing  the large scale shell-model
calculations of low-lying excited states \cite{Langanke99}. Within
this approach a coupling with high-lying resonant states like, e.g.,
the Gamow-Teller resonance, was considered in the framework of the
Brink hypothesis.

The other approach to calculate the rates of weak-interaction
mediated processes at finite temperatures employs the thermal
quasiparticle random phase approximation (TQRPA). For the first time
it was used in \cite{Coop84} to study the electron captures on
neutron-rich nuclei. Recently, there appeared several papers
\cite{Paar09,Dzhioev10,Niu11} where the authors applied TQRPA (or
TRPA) with different nuclear Hamiltonians.

In Ref.\cite{Dzhioev10}, the TQRPA was applied in a framework of a
general context of thermo-field dynamics (TFD) \cite{ume75,ume82} and
with the Hamiltonian consisting of schematic separable effective
interactions. It is well known that schematic separable interactions
allow one to avoid many of computational difficulties and
limitations which are inherent for ``realistic'' residual
interactions. However, the cost is a lack of selfconsistency and
limited predictive power of the calculations for nuclei far from the
valley of stability. Thus, it seems reasonable to use in
calculations for hot nuclei a finite rank separable approximation
elaborated for the residual forces based on an effective interaction
of Skyrme type \cite{Giai98,Sever08,Sever11}. This enables to
combine the advantages of consistency (the mean field and the
residual interaction of RPA are determined from the same effective
interaction) with a computational simplicity (the size of the RPA
problem does not increase with increasing configuration space).

Here we present the formulae which extend the approach of
Refs.~\cite{Giai98} to finite temperatures. The TFD formalism
presented in \cite{ume75,ume82,Ojima81} and adapted to nuclear
structure problems in \cite{Dzhioev09} is used. Having in mind
further application of the formalism to charge-exchange excitations
we consider only the isovector part of the residual particle-hole interaction.

\section{Separabelization of residual interactions}

The exact $p$-$h$ residual interaction $V_{res}$ corresponding to the Skyrme force  can be
obtained as the second derivative of the energy density functional
with respect to the particle density.  In Ref.~\cite{Giai98},
$V_{res}$ was approximated by its Landau-Migdal form.
For the Skyrme
interactions all the Landau parameters with $l>1$ are zero.
Moreover, the Landau parameters with the $l=1$ terms are neglected.
Therefore, the residual interaction in the isovector channel has the
following form:
\begin{equation}
  V_{res}(\mathbf{r}_1,\mathbf{r}_2)=N_0^{-1}[F_0'(r_1)+G'_0(r_1)
  \sigma^{(1)}\sigma^{(2)}]\tau^{(1)}\tau^{(2)}\delta(\mathbf{r}_1-\mathbf{r}_2).
\end{equation}
where $\sigma^{(i)}$ and $\tau^{(i)}$ are the spin and isospin
operators, and $N_0=2k_Fm^*/\pi^2\hbar^2$ with $k_F$ and $m^*$
standing for the Fermi momentum and nucleon effective mass in
nuclear matter. $F'_0$, $G'_0$  are functions of the coordinate
$\mathbf{r}$. Their expressions in terms of the Skyrme parameters
can be found in Ref.~\cite{Giai81}.

Following the method presented in~\cite{Giai98,Sever08} the residual
interaction is reduced to a finite rank separable form
\begin{equation}
   V_{res} = V_M + V_{SM},
\end{equation}
where
\begin{align}\label{V}
  V_M &= -2\sum_{JM}\sum_{n=1}^{N} \kappa^{(n)}_F:\hat M^{(n)\dag}_{JM}\hat M^{(n)}_{JM}:,
  \notag\\
  V_{SM} &= -2\sum_{JM}\sum_{L=J;J\pm1}\sum_{n=1}^{N} \kappa^{(n)}_G:\hat S^{(n)\dag}_{LJM}\hat S^{(n)}_{LJM}:,
\end{align}
and
\begin{equation}
  \binom{\kappa^{(n)}_F}{\kappa^{(n)}_G} = -N_0^{-1}\frac{R\omega_k}{2r_k^2}\binom{F_0'(r_k)}{G'_0(r_k)}.
\end{equation}
Here $R$ is a large enough cutoff radius for a  $N$-point
integration Gauss formula with abscissas $r_k$ and weights
$\omega_k$~\cite{Giai98}.

The operators entering the normal product in Eq.~\eqref{V} are
defined as follows:
\begin{align}
 \hat M^{(n)\dag}_{JM} &= (-1)^{J-M}\hat J^{-1}\sum_{j_nj_pm_nm_p}\langle j_nm_nj_p-m_p|JM\rangle f^{(Jk)}_{j_nj_p}(-1)^{j_p-m_p}a^\dag_{j_nm_n}a_{j_pm_p} ,
  \notag\\
\hat S^{(n)\dag}_{JM} &= (-1)^{J-M}\hat
J^{-1}\sum_{j_nj_pm_nm_p}\langle j_nm_nj_p-m_p|JM\rangle
g^{(LJk)}_{j_nj_p}(-1)^{j_p-m_p}a^\dag_{j_nm_n}a_{j_pm_p},
\end{align}
where $f^{(Jk)}_{j_nj_p}$ is the single-particle matrix elements of
the multipole operators,
\begin{equation}
  f^{(Jk)}_{j_nj_p} = u_{j_n}(r_k)u_{j_p}(r_k)\langle j_n\|i^JY_J\|j_p\rangle,
\end{equation}
and where $g^{(Jk)}_{j_nj_p}$ is the single-particle matrix elements
of the spin-multipole operators,
\begin{equation}
  g^{(LJk)}_{j_nj_p} = u_{j_n}(r_k)u_{j_p}(r_k)\langle j_n\|i^JT_J\|j_p\rangle.
\end{equation}
In the above equations, $\langle j_n\|i^JY_J\|j_p\rangle$  is the
reduced matrix element of the spherical harmonic $Y_{JM}$, $\hat
J=\sqrt{2J+1}$, $T_{LJM}=[Y_J\times\sigma]^J_M$. The radial wave
functions $u_j(r)$ are related to the HF single-particle wave
functions
\begin{equation}
  \phi_{i,m}(1)=\frac{u_i(r_1)}{r_1}\mathcal{Y}^m_{l_i,j_i}(\hat r_1,\sigma_1).
\end{equation}

An interaction in the particle-particle channel has the surface
peaked density-dependent zero-range shape
\begin{equation}\label{pp-channel}
  V_{pair}(\mathbf{r}_1,\mathbf{r}_2)=V_0\left(1 - \eta\left(\frac{\rho(r_1)}{\rho_c}\right)^\alpha\right)\delta(\mathbf{r}_1 - \mathbf{r}_2).
\end{equation}
Here $\rho(r)$ is the particle density in coordinate space, $\rho_0$
is equal  to the nuclear saturation density, $V_0$, $\eta$ and
$\alpha$ are the parameters fixed to reproduce the odd-even mass
difference of nuclei in the study region. The interaction
(\ref{pp-channel}) is responsible for the pairing correlations as
well.

\section{TBCS and TQRPA equations at $\mathbf{T\neq 0}$}

\subsection{Fundamentals of the thermo-field dynamics}

Thermo-field dynamics \cite{ume75,ume82,Ojima81} is a real-time
formalism for treating thermal effects in quantum field theory and
many-body theories. In TFD, the thermal average  of a
given operator $A$ is calculated as the expectation value in
a specially constructed, temperature-dependent state $|0(T)\rangle$
which is termed the thermal vacuum. This expectation value is equal
to the usual grand canonical average of $A$. In this sense, the
thermal vacuum describes the thermal equilibrium of the system.

To construct the state  $|0(T)\rangle$, a formal doubling of the
system degrees of freedom is introduced. In TFD, a tilde conjugate
operator~$\widetilde A$ -- acting in the independent Hilbert space
-- is associated with $A$, in accordance with properly formulated
tilde conjugation rules~\cite{ume75,ume82,Ojima81}
\begin{align}\label{TCR}
 (A_1A_2)\widetilde{} = \widetilde A_1\widetilde A_2,~~
 (c_1A_1&+ c_2A_2)\widetilde{} = c^*_1 \widetilde A_1 + c^*_2 \widetilde A_2,
  \notag\\
  (A^\dag)\widetilde{}=(\widetilde A)^\dag,&~~(\widetilde A)\widetilde{} = A,
\end{align}
where $A_1$ and $A_2$ stand for any operators and $c_1$ and $c_2$ are $c$-numbers. The asterisk denotes the complex conjugate.
It is assumed that any bosonic (fermionic) operator with tilde commutes (anticommutes)
with all bosonic (fermionic) operators without tilde.

Let $H$ be the Hamiltonian of the system. In the doubled Hilbert space
the thermal vacuum is defined as the zero-energy eigenstate of the so-called thermal
Hamiltonian $\mathcal H=H-\widetilde H$ and it satisfies the thermal state
condition~\cite{ume75,ume82,Ojima81}
\begin{equation}\label{TSC}
A|0(T)\rangle = \sigma\,{\rm e}^{{\mathcal H}/2T} {\widetilde
A}^\dag|0(T)\rangle,
\end{equation}
where  $\sigma=1$ for bosonic~$A$ and $\sigma=i$ for fermionic $A$.
The thermal state condition is one of the most fundamental
relations in TFD. The celebrated Kubo-Martin-
Schwinger condition, which is one of the basic axioms in
the $c^*$ -algebra formalism for statistical mechanics, is a result
of the thermal state condition in TFD. Furthermore, from~\eqref{TSC}
we can see immediately that, in TFD, there always exists
a certain combination of $A$ and $\widetilde A^\dag$ which annihilates
the thermal vacuum.  That mixing is promoted by a specific canonical transformation called the thermal Bogoliubov
transformation~\cite{ume75,ume82}. The
temperature dependence comes from the transformation parameters.

The important point is that in the doubled Hilbert
space the time-translation operator is the thermal
Hamiltonian $\mathcal H=H-\widetilde H$. This means that the excitations of the
thermal system are obtained by the diagonalization of
$\mathcal H$. The existence of the thermal vacuum annihilation
operators allows for straightforward extensions of different zero-temperature approximations to diagonalize the thermal Hamiltonian.
As  follows from the definition of $\mathcal H$ each of its
eigenstates with positive energy has the counterpart -- the
tilde-conjugate eigenstate -- with negative but the
same absolute energy value. Transitions from the thermal vacuum to positive (non-tilde) energy states correspond to
excitation of the system, while transitions to negative (tilde) energy states - to deexcitation.

\subsection{Equations for pairing correlations at $\mathbf{T\neq0}$}

Applying the TFD formalism we suppose that the nuclear proton and
neutron mean fields are already produced according the Hartree-Fock
procedure and our Hamiltonian $H$ consists of the mean fields and
separabelized residual interactions presented in the previous
Section~2. In particular, it means that we ignore the influence of
temperature on the nuclear mean field.

At first, following the TFD prescription, we double the original
nuclear degrees of freedom introducing the so-called tilde creation
and annihilation operators $\widetilde{a}^\dag_{jm},
\widetilde{a}_{jm}$ and construct the thermal Hamiltonian
\[
\mathcal{H} = H - \widetilde{H}
\]
Then we transform $\mathcal{H}$ to the thermal quasiparticle
representation by means of the two canonical transformations. The
first one is the standard Bogoliubov transformation to quasiparticle
operators
 \begin{align}\label{B_tr}
   a^\dag_{jm}&=u_j\alpha^\dag_{jm} + v_j\alpha_{\overline{jm}} \nonumber \\
\widetilde{a}^\dag_{jm}&=u_j\widetilde{\alpha}^\dag_{jm} +
v_j\widetilde{\alpha}_{\overline{jm}},~~ (u^2_j+v^2_j=1),
 \end{align}
where $\alpha_{\overline{jm}} =(-1)^{j-m}\alpha_{j-m}$. The second transformation is the thermal Bogoliubov
transformation~\cite{ume75,ume82}. It mixes the quasiparticle and
tilde quasiparticle operators, thus producing thermal quasiparticle
operators and their tilde partners
$\beta^\dag_{jm},~\beta^{\phantom{\dag}}_{jm},~
\widetilde\beta^\dag_{jm},~\widetilde\beta^{\phantom{\dag}}_{jm}$
\begin{align}\label{T_tr}
  \beta^\dag_{jm}&=x_j\alpha^\dag_{jm}-i y_j\widetilde\alpha_{jm}~, \nonumber\\
  \widetilde\beta^\dag_{jm}&=x_j\widetilde\alpha^\dag_{jm}+i
  y_j\alpha_{jm}~,~~
  (x^2_j+y^2_j=1)~.
\end{align}

The coefficients of both the transformations are determined by the
diagonalization of the sum of single-particle and pairing parts of
$\mathcal{H}$ and additional demand for the BCS thermal vacuum to
obey the thermal state condition \cite{Dzhioev09}. At the end, we
get the following equations for the coefficient $u_j, v_j$ and $x_j,
y_j$:
\begin{eqnarray}
v_j & = & \frac{1}{\sqrt
2}\left(1-\frac{E_j-\lambda}{\varepsilon_j}\right)^{1/2},\
 u_j=(1-v_j^2)^{1/2}, \label{u&v} \\
y_j & = &
\left[1+\exp\left(\frac{\varepsilon_j}{T}\right)\right]^{-1/2},\
  x_j=\bigl(1-y^2_j\bigr)^{1/2}, \label{x&y}
\end{eqnarray}
where $\varepsilon_j=\sqrt{(E_j-\lambda)^2+\Delta^2_{j}}$ is a
quasiparticle energy. The pairing gaps $\Delta_j$  and the chemical
potential $\lambda$ are the solutions to the finite-temperature BCS
equations
\begin{align}\label{BCS}
\Delta_{j}(T)&=-\sum_{j\,'}(-1)^{l_j+l_{j\,'}}
\sqrt{\frac{2j\,'+1}{2j+1}}G_0(jjj\,'j\,')(1-2y^2_{j\,'})u_{j\,'}v_{j\,'},\nonumber\\
N&=\sum_j(2j+1)(v^2_jx^2_j+u^2_jy^2_j),
\end{align}
where $N$ is the number of neutrons or protons in a nucleus, and
two-body matrix element $G_0(jjj\,'j\,')$ is given by the $J=0$
particle-particle matrix element of the interaction $V_{pair}$
(\ref{pp-channel})
\begin{equation}
  G_0(jjj\,'j\,') = \langle jj| V_{pair}| j\,'j\,'\rangle_{00}.
\end{equation}

Now, the sum of the single-particle and the pairing parts of the
Hamiltonian $\mathcal H$ becomes diagonal
\[
\mathcal{H}_\text{BCS} = \sum_{\tau=\mathrm{p,n}}{\sum_{jm}}^\tau \varepsilon_{j} \left(
\beta^\dag_{jm} \beta^{\phantom{\dag}}_{jm} -
\widetilde\beta^\dag_{jm} \widetilde\beta^{\phantom{\dag}}_{jm}
\right).
\]
Here ${\sum}^\tau$ implies a summation over proton or neutron single-particle states only.
The Hamiltonian $\mathcal{H}_\text{BCS}$  describes a
system of noninteracting non-tilde and tilde thermal quasiparticles  with energies
$\varepsilon_{j}$ and  $-\varepsilon_{j}$, respectively. The vacuum of thermal quasiparticles, $|0(T);\mathrm{qp}\rangle$, is the thermal
vacuum in BCS approximation.

\subsection{Thermal QRPA equations}

The transformations (\ref{B_tr}) and (\ref{T_tr}) are applied to the
whole thermal nuclear Hamiltonian $\mathcal{H}$. At the next step we
approximately diagonalize $\mathcal{H}$ within the Thermal Quasiparticle Random Phase Approximation
\begin{equation}\label{H_TQRPA}
  {\cal H}\approx {\cal H}_{\mathrm{TQRPA}}=\sum_{JMi}\omega_{J i}(Q^\dag_{JM i}Q_{JM i} - \widetilde Q^\dag_{JM i} \widetilde Q_{JM i}).
\end{equation}
Here the thermal (charge-exchange) phonon creation operator is defined as a linear superposition of the proton-neutron thermal quasiparticle pair creation and
annihilation operators
\begin{multline}\label{ch_phonon}
 Q^\dag_{JM i}=\sum_{j_n j_p}
  \Bigl(
  \psi^{J i}_{j_nj_p}[\beta^\dag_{j_n}\beta^\dag_{j_p}]^J_M+
  \widetilde\psi^{J i}_{j_nj_p}[\widetilde\beta^\dag_{\overline{\jmath_n}}
  \widetilde\beta^\dag_{\overline{\jmath_p}}]^J_M+
  i\eta^{J i}_{j_n j_p}[\beta^\dag_{j_n} \widetilde\beta^\dag_{\overline{\jmath_p}}]^J_M+
  i\widetilde\eta^{J i}_{j_nj_p}[\widetilde\beta^\dag_{\overline{\jmath_n}} \beta^\dag_{j_p}]^J_M
  \\
  +
   \phi^{J i}_{j_nj_p}[\beta_{\overline{\jmath_n}}\beta_{\overline{\jmath_p}}]^J_M+
   \widetilde\phi^{J i}_{j_n j_p}[\widetilde\beta_{j_n}\widetilde\beta_{j_p}]^J_M+
   i\xi^{J i}_{j_n j_p}[\beta_{\overline{\jmath_n}}\widetilde\beta_{j_p}]^J_M+
   i\widetilde\xi^{J i}_{j_nj_p}[\widetilde\beta_{j_n} \beta_{\overline{\jmath_p}}]^J_M
   \Bigr),
\end{multline}
and $[~]^J_M$ denotes the coupling of single-particle angular
momenta $j_n, j_p$ to total angular momentum $J$. The tilde conjugate thermal phonon
operator  $\widetilde Q^\dag_{JM i}$ can be obtained
from~\eqref{ch_phonon} by applying tilde conjugation rules~\eqref{TCR}. 
Now the thermal equilibrium
state is treated as the vacuum $|0(T);{\rm
ph}\rangle$ for thermal
phonon annihilation operators and it obeys the thermal state condition~\eqref{TSC}.  The excited thermal one-phonon states are
$Q^\dag_{JM i}|0(T);{\rm ph}\rangle$ and $\widetilde
Q^\dag_{\overline{JM} i}|0(T);{\rm ph}\rangle$.

The thermal phonon operators are assumed to commute as bosonic
operators, that is, ${[Q^{\phantom\dag}_{JM i},Q^{\dag}_{J'M'
i'}]=\delta_{JJ'}\delta_{MM'}\delta_{ii'}}$. This assumption imposes
the normalization condition on the phonon amplitudes
 \begin{multline}\label{constraint}
 \sum_{j_n j_p}\Bigl(
  \psi^{J i}_{j_nj_p}\psi^{J i'}_{j_nj_p}+
 \widetilde\psi^{J i}_{j_nj_p}\widetilde\psi^{J i'}_{j_nj_p}+
  \eta^{J i}_{j_nj_p}\eta^{J i'}_{j_nj_p}+
  \widetilde\eta^{J i}_{j_nj_p}\widetilde\eta^{J i'}_{j_nj_p}\\
  -\phi^{J i}_{j_nj_p}\phi^{J i'}_{j_pj_n}-
  \widetilde\phi^{J i}_{j_pj_n}\widetilde\phi^{J i'}_{j_nj_p}-
   \xi^{J i}_{j_nj_p} \xi^{J i'}_{j_nj_p}-
  \widetilde\xi^{J i}_{j_nj_p}\widetilde\xi^{Ji'}_{j_nj_p}\Bigr)=
      \delta_{ii'}.
  \end{multline}
Furthermore, additional constraints on the amplitudes come from the thermal state condition. Namely,  putting $A=[\alpha^\dag_{j_n} \alpha^\dag_{j_p}]^J_M$ in~\eqref{TSC} 
we get ($\omega_{Ji} >0$)
\begin{align}\label{constr1}
  (x_{j_n}x_{j_p} \psi^{J i}_{j_nj_p} + y_{j_n} y_{j_p}\widetilde\phi^{J i}_{j_nj_p}) &=
  \exp\Bigl(\frac{\omega_{Ji}}{2T}\Bigr) (x_{j_n}x_{j_p} \widetilde\phi^{J i}_{j_nj_p}  + y_{j_n} y_{j_p}\psi^{J i}_{j_nj_p}),
   \notag \\
  (x_{j_n}x_{j_p}\widetilde \psi^{J i}_{j_nj_p} + y_{j_n} y_{j_p}\phi^{J i}_{j_nj_p}) &=
  \exp\Bigl(-\frac{\omega_{Ji}}{2T}\Bigr) (x_{j_n}x_{j_p}\phi^{J i}_{j_nj_p}  + y_{j_n} y_{j_p} \widetilde\psi^{J i}_{j_nj_p}),
\end{align}
and for $A=[\alpha^\dag_{j_n} \alpha_{\overline{j_p}}]^J_M$ we have
 \begin{align}\label{constr2}
  (x_{j_n}y_{j_p} \eta^{J i}_{j_nj_p} + y_{j_n} x_{j_p}\widetilde\xi^{J i}_{j_nj_p}) &=
  \exp\Bigl(\frac{\omega_{Ji}}{2T}\Bigr) (x_{j_n}y_{j_p} \widetilde\xi^{J i}_{j_nj_p}  + y_{j_n} x_{j_p}\eta^{J i}_{j_nj_p}),
   \notag \\
  (x_{j_n}y_{j_p}\widetilde \eta^{J i}_{j_nj_p} + y_{j_n} x_{j_p}\xi^{J i}_{j_nj_p}) &=
  \exp\Bigl(-\frac{\omega_{Ji}}{2T}\Bigr) (x_{j_n}y_{j_p}\xi^{J i}_{j_nj_p}  + y_{j_n} x_{j_p} \widetilde\eta^{J i}_{j_nj_p}).
\end{align}

To find the energy and the structure of thermal phonons we apply the variational principle, i.e. we minimize
the expectation value of ${\cal H}$ over the thermal one-phonon state under the
constraints~\eqref{constraint}. As a result we get the TQRPA eigenvalue equations for the amplitudes and
the energy eigenvalues $\omega_{Ji}$.
In contrast to the zero-temperature case, the negative eigenvalues of TQRPA matrix have physical meaning 
and they can be interpreted as the excitation energies of tilde thermal one-phonon states relative to the thermal vacuum.
Besides, each eigenvalue is twice degenerate so that ${\cal H}_{\mathrm{TQRPA}}$~\eqref{H_TQRPA} is invariant
under the thermal Bogoliubov transformation
\begin{equation}
  Q^\dag_{JMi}\to X_{Ji} Q^\dag_{JMi} -  Y_{Ji}\widetilde Q_{JMi},~~~  \widetilde Q^\dag_{JMi}\to X_{Ji}\widetilde Q^\dag_{JMi} -  Y_{Ji} Q_{JMi}
\end{equation}
with $X^2_{Ji} - Y^2_{Ji}=1$. To find the structure of thermal phonons unambiguously we demand that the constraints~(\ref{constr1},\ref{constr2}) are valid. 
Only in this case the vacuum of thermal phonons is the thermal vacuum in the TQRPA approximation.

Using the separable form
of the residual interaction one can reduce remarkably the dimensions
of the corresponding TQRPA matrixes. To do this we introduce a
vector $\binom{D_+}{D_-}$ by its components:
 \begin{align}
 D^{Jin}_{\pm} = \sum_{j_nj_p} d^{(Jn)}_{j_nj_p}\Bigl\{
 u^{(\pm)}_{j_nj_p}\bigl[x_{j_n}x_{j_p}(\psi^{J i}_{j_nj_p}\pm\phi^{J i}_{j_nj_p})\pm y_{j_n}y_{j_p}(\widetilde\psi^{J i}_{j_nj_p}\pm\widetilde\phi^{J i}_{j_nj_p})\bigr] +
 \notag \\
 +v^{(\mp)}_{j_nj_p}\bigl[x_{j_n}y_{j_p}(\eta^{J i}_{j_nj_p}\pm\xi^{J i}_{j_nj_p})\pm y_{j_n}x_{j_p}(\widetilde\eta^{J i}_{j_nj_p}\pm\widetilde\xi^{J i}_{j_nj_p})\bigr]
  \Bigr\},
  \end{align}
where $u^{(\pm)}_{j_nj_p} = u_{j_n}v_{j_p}\pm v_{j_n}u_{j_p}$ and
$v^{(\pm)}_{j_nj_p} = u_{j_n}u_{j_p}\pm v_{j_n}v_{j_p}$.  Phonon
amplitudes are functions of the vectors
\begin{multline}\label{amplitude}
\binom{\psi}{\phi}^{J i}_{j_nj_p}=
  \frac{\hat J^{-2}}{\varepsilon^{(+)}_{j_nj_p}\mp\omega_{J i}}\bigl(x_{j_n}x_{j_p}X_{\lambda i}-y_{j_n}y_{j_p}Y_{\lambda i}\bigr)\sum_{n=1}^{2N}
  d^{(Jn)}_{j_nj_p} \kappa^{(n)}\Bigl( D^{Jin}_{+} u^{(+)}_{j_nj_p} \pm  D^{Jin}_{-} u^{(-)}_{j_nj_p}\Bigr),\\
\shoveleft{\binom{\widetilde\psi}{\widetilde\phi}^{J i}_{j_nj_p}}=
  \frac{\hat J^{-2}}{\varepsilon^{(+)}_{j_nj_p}\pm\omega_{J i}}\bigl(x_{j_n}x_{j_p}Y_{\lambda i}-y_{j_n}y_{j_p}X_{\lambda i}\bigr)\sum_{n=1}^{2N}
  d^{(Jn)}_{j_nj_p} \kappa^{(n)}\Bigl( D^{Jin}_{+} u^{(+)}_{j_nj_p} \mp  D^{Jin}_{-} u^{(-)}_{j_nj_p}\Bigr),\\
\shoveleft{\binom{\eta}{\xi}^{J i}_{j_nj_p}}=
  \frac{\hat J^{-2}}{\varepsilon^{(-)}_{j_nj_p}\mp\omega_{J i}}\bigl(x_{j_n}y_{j_p}X_{\lambda i}-y_{j_n}x_{j_p}Y_{\lambda i}\bigr)\sum_{n=1}^{2N}
  d^{(Jn)}_{j_nj_p} \kappa^{(n)}\Bigl( D^{Jin}_{+} v^{(-)}_{j_nj_p} \pm  D^{Jin}_{-} v^{(+)}_{j_nj_p}\Bigr),\\
\shoveleft{\binom{\widetilde\eta}{\widetilde\xi}^{J i}_{j_nj_p}}=
  \frac{\hat J^{-2}}{\varepsilon^{(-)}_{j_nj_p}\pm\omega_{J i}}\bigl(x_{j_n}y_{j_p}Y_{\lambda i}-y_{j_n}x_{j_p}X_{\lambda i}\bigr)\sum_{n=1}^{2N}
  d^{(Jn)}_{j_nj_p} \kappa^{(n)}\Bigl( D^{Jin}_{+} v^{(-)}_{j_nj_p} \mp  D^{Jin}_{-} v^{(+)}_{j_nj_p}\Bigr),
\end{multline}
where
\begin{equation}
Y_{J i}=\left[\exp\left(\frac{\omega_{J i}}{T}\right)-1\right]^{-1/2},~~~ X_{Ji}=(1+Y^2_{\lambda i})^{1/2}.
\end{equation}

So, the TQRPA equations are reduced to the set of equations for $D^{Jin}_{\pm}$:
\begin{equation}
  \left(\begin{array}{cc}
     \mathcal{M}_1-\frac12I & \mathcal{M}_2 \\
     \mathcal{M}_2 & \mathcal{M}_3-\frac12I
   \end{array}\right)\left(
   \begin{array}{c}
     D_+ \\  D_-
   \end{array}\right)=0.
\end{equation}
The matrix elements of the $2N\times2N$ matrices $\mathcal{M}_\beta$ have the following expressions
\begin{eqnarray*}
\mathcal{M}^{nn'}_{1,3}&=&\frac{\kappa^{(n')}}{\hat J^2}\!\sum_{j_n j_p}
   d^{(Jn)}_{j_n j_p}d^{(Jn')}_{j_n j_p}\left\{
      \frac{\varepsilon_{j_nj_p}^{(+)}(u^{(\pm)}_{j_nj_p})^2}
       {(\varepsilon_{j_nj_p}^{(+)})^2-\omega^2_{Ji}}(1\!-\!y^2_{j_n}\!-\!y^2_{j_p})-
      \frac{\varepsilon_{j_nj_p}^{(-)}(v^{(\mp)}_{j_nj_p})^2}
       {(\varepsilon_{j_nj_p}^{(-)})^2-\omega^2_{Ji}}(y^2_{j_n}\!-\!y^2_{j_p})\right\},\\
\mathcal{M}^{nn'}_{2}&=&\frac{\kappa^{(n')}}{\hat J^2}\,\omega_{Ji}\sum_{j_n j_p}
  d^{(Jn)}_{j_n j_p}d^{(Jn')}_{j_n j_p}\left\{
  \frac{u^{(+)}_{j_nj_p}u^{(-)}_{j_nj_p}}
       {(\varepsilon_{j_nj_p}^{(+)})^2-\omega^2_{Ji}}(1-y^2_{j_n}-y^2_{j_p})-
  \frac{v^{(+)}_{j_nj_p}v^{(-)}_{j_nj_p}}
       {(\varepsilon_{j_nj_p}^{(-)})^2-\omega^2_{Ji}}(y^2_{j_n}-y^2_{j_p})\right\},
\end{eqnarray*}
where $\varepsilon_{j_nj_p}^{(\pm)} = \varepsilon_{j_n} \pm \varepsilon_{j_p}$.
Its solution requires to compute the determinant
   \begin{equation}
 \mathrm{det} \left(\begin{array}{cc}
     \mathcal{M}_1-\frac12I & \mathcal{M}_2 \\
     \mathcal{M}_2 & \mathcal{M}_3-\frac12I
   \end{array}\right)=0
\end{equation}
 and we find the eigenvalues of the TQRPA equations.

\section{Charge-exchange transition probabilities}

Charge-exchange transition probabilities  (transition strengths)
from the thermal vacuum to thermal one-phonon states are given by
the squared reduced matrix elements of the corresponding transition
operator:
\begin{align}\label{trans_ampl}
\Phi^{(\pm)}_{J i}&=
 \left|\langle Q_{JM i}  \|D^{(\pm)}_{JM}\|0(T);\mathrm{ph}\rangle\right|^2,
    \notag\\
\widetilde \Phi^{(\pm)}_{\lambda i}&= \left|\langle \widetilde Q_{\overline{JM} i}  \|D^{(\pm)}_{JM}\|
0(T);\mathrm{ph}\rangle\right|^2.
\end{align}
Hereinafter the symbol $(-)$ labels the $n\to p$ transition
operators, and the symbol $(+)$ labels the $p\to n$ transition
operators. The explicit expressions for $\Phi^{(\pm)}_{\lambda i}$
and $\widetilde \Phi^{(\pm)}_{\lambda i}$ are the following:
\begin{eqnarray}\label{ph_tr_ampl}
\Phi_{J i}^{(+)}&=&\Bigl(\sum_{j_nj_p} d^{(+)}_J(j_nj_p)\, \Omega_1(j_nj_p;J
i)\Bigr)^2,\notag
   \\
\widetilde\Phi_{J i}^{(+)}&=&\Bigl(\sum_{j_nj_p} d^{(+)}_J(j_nj_p)\,
\widetilde\Omega_1(j_nj_p;J i)\Bigr)^2,\notag
   \\
\Phi_{J i}^{(-)}&=&\Bigl(\sum_{j_nj_p}(-1)^{j_p-j_n+J}
d^{(-)}_J(j_pj_n)\,\Omega_2(j_nj_p;J i)\Bigr)^2,\notag
   \\
\widetilde\Phi_{J i}^{(-)}&=&\Bigl(\sum_{j_nj_p}(-1)^{j_p-j_n+J}
d^{(-)}_J(j_pj_n)\,\widetilde\Omega_2(j_nj_p;J i)\Bigr)^2,
\end{eqnarray}
where~$d^{(\mp)}_\lambda(j_{p(n)}j_{n(p)})$ is a reduced single-particle matrix element of the
transition operator
\begin{equation}
d^{(\mp)}_\lambda(j_{p(n)}j_{n(p)})=\langle j_{n(p)}\|D^{(\mp)}_{\lambda}\|j_{p(n)}\rangle,
\end{equation}
functions $\Omega_{1,2}(j_nj_p;J i)$  are linear combinations of the phonon
amplitudes~\eqref{amplitude}:
\begin{align}\label{f1}
\Omega_1(j_nj_p;J i)&=
   u_{j_n}v_{j_p}\!\bigl(x_{j_n}x_{j_p}\psi^{J i}_{j_nj_p}\!+\!
                       y_{j_n}y_{j_p}\widetilde\phi^{J i}_{j_nj_p}\bigr)
             \!+\!
   v_{j_n}u_{j_p}\bigl(y_{j_n}y_{j_p}\widetilde\psi^{J i}_{j_nj_p}\!+\!
                       x_{j_n}x_{j_p}\phi^{J i}_{j_nj_p}\bigr)+ \nonumber\\
                       &+\!
   u_{j_n}u_{j_p}\!\bigl(x_{j_n}y_{j_p}\eta^{J i}_{j_nj_p}\!+\!
                       y_{j_n}x_{j_p}\widetilde\xi^{J i}_{j_nj_p}\bigr)
             \!-\!
   v_{j_n}v_{j_p}\!\bigl(y_{j_n}x_{j_p}\widetilde\eta^{J i}_{j_nj_p}\!+\!
                        x_{j_n}y_{j_p}\xi^{J i}_{j_nj_p}\bigr),\nonumber\\
                                              \nonumber\\
\Omega_2(j_nj_p;\lambda i)&=
    v_{j_n}u_{j_p}\!\bigl(x_{j_n}x_{j_p}\psi^{J i}_{j_nj_p}\!+\!
                       y_{j_n}y_{j_p}\widetilde\phi^{J i}_{j_nj_p}\bigr)
             \!+\!
   u_{j_n}v_{j_p}\bigl(y_{j_n}y_{j_p}\widetilde\psi^{J i}_{j_nj_p}\!+\!
                       x_{j_n}x_{j_p}\phi^{J i}_{j_nj_p}\bigr)- \nonumber\\
                       &+\!
   v_{j_n}v_{j_p}\!\bigl(x_{j_n}y_{j_p}\eta^{J i}_{j_nj_p}\!+\!
                       y_{j_n}x_{j_p}\widetilde\xi^{J i}_{j_nj_p}\bigr)
             \!+\!
   u_{j_n}u_{j_p}\!\bigl(y_{j_n}x_{j_p}\widetilde\eta^{J i}_{j_nj_p}\!+\!
                        x_{j_n}y_{j_p}\xi^{J i}_{j_nj_p}\bigr),
\end{align}
and $\widetilde\Omega_{1,2}(j_nj_p;J i)$ result from  $\Omega_{1,2}(j_nj_p;J i)$ by
changing non-tilde phonon amplitudes by their tilde partners and vise versa.

The excitation energies with respect to the thermal equilibrium
state of the parent nucleus are
\begin{equation}\label{E1}
  E^{\mp} = \omega_{Ji}\mp\Delta\lambda
\end{equation}
for non-tilde phonon states, and
\begin{equation}\label{E2}
  E^{\mp} = -\omega_{Ji}\mp\Delta\lambda
\end{equation}
for tilde phonon states. Here $\Delta\lambda=\lambda_n-\lambda_p$ is the difference between the neutron and the proton chemical potentials.

Expressions~\eqref{ph_tr_ampl} and (\ref{E1}, \ref{E2}) determine charge exchange
strength distribution in a hot nucleus within the TQRPA. Note that at finite temperatures
some amount of transition strength is always located
in the region of negative transition energies.

\section*{Acknowledgements}

The authors are grateful to Dr. A.\,Severyukhin, Prof. E. Khan and
participants of the seminar of Theory Group of the Institut Physique
Nuclea\'ire (Orsay) for valuable discussions and comments. The
authors also thank the hospitality of IPN (Orsay) where the main
part of the work was done. This work was supported by the grant of
CNRS-RFBR 11-091054.

\end{document}